\documentclass{amsart}

\usepackage{amssymb}
\newcommand{\W}{\mathcal{W}_{D+p}}
\newcommand{\M}{\mathcal{M}_{D}}
\newcommand{\K}{\mathcal{K}_{p}}
\newcommand{\pd}{\partial}
\newcommand{\bl}{\bar{\lambda}}
\newcommand{\tr}{\mathop{\rm tr}}
\newcommand{\Res}{\mathop{\rm Res}}
\newcommand{\vol}{\mathop{\rm vol}}
\newcommand{\R}{\mathop{\rm Re}}
\begin{document}

\title[Life beyond...]{Compactified Quantum Fields. Is there
Life Beyond the Cut-off Scale?}
\author{Corneliu Sochichiu}%
\address{Institutul de Fizic\u a Aplicat\u a A\c S \\
str. Academiei, nr. 5, Chi\c sin\u au MD2028 \\
MOLDOVA}%
\address{Bogoliubov Laboratory of Theoretical Physics\\
Joint Institute for Nuclear Research\\ 141980 Dubna, Moscow Reg.\\
RUSSIA}
\email{sochichi@thsun1.jinr.ru}%

\thanks{Work supported by RFBR grant \# {99-01-00190}, INTAS
grant \# {95-0681}, Scientific
School support grant \# {96-15-06208} and HLP 99-10}%

\begin{abstract}
A consistent definition of high dimensional compactified quantum
field theory without breaking the Kaluza--Klein tower is proposed.
It is possible in the limit when the size of compact dimensions is
of the order of the cut off. This limit is nontrivial and depends
on the geometry of compact dimensions. Possible consequences are
discussed for the scalar model.
\end{abstract}
\maketitle
\section*{Introduction}
The Nature at high energies is supposed to be described by some
string models (see e.g. \cite{gsw}). In the local approximation
the strings are known to be described at tree level field models.
Strings are formulated in high-dimensional space-time, therefore
the respective field models are also high-dimensional. The low
energy observable world, however, is low-dimensional and is
described by local \emph{quantum} field theories.
\emph{Compactification} provides a mechanism that lower the
effective dimensionality of the model in low energy limit. In 80's
it was shown (see \cite{witten}) that at classical level
compactification may lead to low-dimensional ($D=4$) unified
models through Kaluza--Klein mechanism.

Thus, one is justified to expect the compactification to appear
due to quantum corrections to the high-dimensional field model.
The main problem the quantum treatment is faced with is one of
\emph{non-renormalisability} proper for the high-dimensional field
models.

The possible one-loop counterterms arising in high-dimensional
quantum field models were computed many years ago \cite{toms}, and
are known indeed to lead to non-renormalisability of the effective
action at any \emph{finite} size of the compact dimensions. The
model become worse as the number of compact dimensions increase.

The non-renormalisability in high dimensions is an UV feature and
it persists also in compactified models (unless the
compactification size reach the UV region!).

A way to avoid the problem with non-renormalisability would be
consideration of a larger model, i.e. of the string
theory\footnote{In Refs. \cite{anton}, the problem of the
supersymmetry  breaking and coupling renormalisation induced by
compactification was studied in the framework of string theory. It
was shown that for some compactifications the the low dimensional
models feel extra dimensions at energies $\sim 1~TeV$.} in the
case when the high dimensional model under consideration is the
effective field theory for such a string model, in this case to
absorb the non-renormalisable divergences one has to consider
nonlocal modes, thus leaving the framework of local field theory.

The careful analysis here can, probably, be done but the field
theoretical approach is simpler and, we believe, more instructive.
We believe that the quantum field theory approach starts to be
applicable when the size of the compact dimensions is close enough
to zero. Quantitatively this may happen when the size of compact
dimensions is of the order of UV cutoff (or even higher), i.e.
when the contribution of the order of size of compact dimensions
is negligible.

Compactification can be realised in different ways the simplest
one being when the high-dimensional space-time is represented as a
product of a compact space with (low-dimensional) Minkowski
space-time. Compactification to circle of the scalar field was
considered in our previous work \cite{eu}.

Spectrum of compactified models consists of propagating part which
correspond to modes which are ``constant'' in the compact
directions, they form the field content of the low-dimensional
model. Beyond the propagating part there is an infinite number of
massive Kaluza--Klein (KK) modes. Their masses are of the order of
inverse specific size of the compact dimensions. When this size is
small, the respective masses are large and KK modes cannot
propagate in low dimensions, but due to interaction with
propagating modes their quantum fluctuations can influence the
dynamics of the latter even at low energies.

The aim of the actual paper is to investigate which is the
condition of existence of compactified model as a renormalisable
low-dimensional effective quantum field theory, and which is the
effect of the high-dimensions to this theory. These days
phenomenological aspects of the compactified high dimensional
models are intensively studied in the framework of GUT building,
(for a review see e.g. \cite{ddg}), although the nature of
compactified models as QFT in these investigations is not yet
clear.
 We hope that actual study clarifies some  of these points.

Our aim is integration of KK modes in order to get the low
dimensional effective action for remaining propagating fields. The
non-renormalisability problem here takes the form of divergence of
the sums over the KK contributions in some Feynman diagrams
although the loop contribution of each KK field corresponding to
this diagram taken separately is renormalisable and vanishes due
in the limit of the small compactification size. These
divergencies, however, are regularised by the same UV
regularisation we use to compute each KK diagram. Combining cutoff
removing with the zero limit of the compactification size one may
get some finite contribution to effective action from the KK
modes. It may seem that in this limit the model is trivially
reduced to a low dimensional one which ``forgot'' everything about
the higher dimensions. We show that even there the low dimensional
effective model has strong dependence on the geometry of compact
dimensions.

Such a combined limiting procedure means that the (physical) size
of compact dimensions is in UV region. In other words the validity
of QFT approach is limited to energies below the inverse
compactification size. Also, we consider that at energies where
the QFT approach is applicable gravity is in its classical regime.
Otherwise, one is thrown from the local QFT approach due to
non-renormalisability of gravitational interactions.

The plan of the paper is as follows. We consider scalar model with
$\frac{\lambda}{4!}\phi^4$ interaction in a high-dimensional
space-time. First we compactify the space-time on the product of
low-dimensional Minkowski space-time and a compact Riemannian
manifold. After that we integrate over KK and high energy modes of
propagating field i.e. find their one loop-contribution. Then we
find the renormalisability conditions for obtained model, and,
finally analyse the dependence of the low dimensional model on the
compactification geometry.

\section{Compactified Model Renormalisation}

In what follows we will consider the model of selfinteracting
scalar field $\phi$ on a $(D+p)$-dimensional space-time manifold
$\W$. It is described by the classical action,
\begin{equation}\label{action}
  S_{(D+p)}=\int_{\W}d^{D+p}x \left\{\frac12 \pd_M\phi \pd^M\phi -\frac12
  m^2 \phi^2 -V(\phi)\right\},
\end{equation}
where capital roman letters $M,N=0,1,\dots,D+p-1$, are associate
with indices of high-dimensional space-time manifold, metric
signature is chosen to be $+--\dots -$. We assume potential
$V(\phi)$ to be renormalisable in $D$-dimensions. Thus if one
compactifies this model to $D=4$ the potential $V(\phi)$ should be
at most quartic in $\phi$. For definiteness we will  consider
that,
\begin{equation}\label{potential}
  V(\phi)=\frac{\lambda}{4!}\phi^4.
\end{equation}

Let the manifold $\W$ on which the fields are defined be
represented as a product $\W=\M\times\K$, where the $\M$ is
$D$-dimensional Minkowski space-time and $\K$ is a compact
Riemannian manifold. In what follows greek indices $\mu,\nu,
\dots$ enumerate coordinates $x^\mu$ of $\M $ and latin indices
$i,j,\dots$, respectively, coordinates $y^i$ of $\K$.

Kinetic term of the action (\ref{action}) can be split in the
Minkowski and $\K$-part according to the decomposition
$\W=\M\times\K$ as follows,
\begin{equation}\label{split}
  \frac12 \pd_M\phi
  \pd^M\phi=\frac12\pd_{\mu}\phi\pd^{\mu}\phi+\frac12\phi\Delta\phi
\end{equation}
where $\frac12\pd_{\mu}\phi\pd^{\mu}\phi$ is kinetic term on the
Minkowski space-time $\M$, and
\begin{equation}\label{lap}
  \Delta=\frac{1}{\sqrt{\det
g}}\pd_{i}g^{ij}\sqrt{\det g}\pd_{j},
\end{equation}
is the scalar Laplace operator on $\K$.

Since $\K$ is compact the operator $\Delta$ has discrete spectrum
with orthonormal eigenfunctions
\begin{align}\label{spec1}
  \Delta \psi_n= -\mu_n^2 \psi_n,\\ \label{spec2}
  \int_{\K}dy\, \psi_n^* \psi_m =\delta_{nm},
\end{align}
where $dy\equiv \sqrt{\det (g)}d^py$ is the invariant measure on
$\K$ and $n,m$ span a $p$-di\-men\-sio\-nal (generally irregular)
lattice $\Gamma$. Complex conjugate of an eigenfunction is an
eigenfunction again,
\begin{equation}\label{real}
  \psi_{n}^{*}=\psi_{n^{*}},
\end{equation}
with $\mu_{n^{*}}^2=\mu_{n}^2$.

Let us note that zero modes of $\Delta$ are locally constant
functions. The number of independent locally constant functions is
equal to the number of connected components of $\K$. This
represents well known fact from the cohomology theory (see e.g.
\cite{kn}), as well as from the properties of scalar Laplace operator.
In any case, eigenfunctions of $\Delta$ can be chosen in such a
way to have support on a single connected component of $\K$.

One can decompose $\phi(x,y)$ in terms of eigenfunctions of
$\Delta$,
\begin{align}\label{decomp}
  \phi(x,y)=\sum_{n\in\Gamma} \phi_n (x)\psi_n (y),\\ \text{where,} \qquad
  \phi_n (x)= \int_{\K} dy\, \psi^*(y)\cdot \phi(x,y).
\end{align}

In terms of this decomposition the classical action (\ref{action})
looks as follows
\begin{equation}\label{d-p}
  S_D=\int_{\M}dx\left\{\frac12 \sum_n(\pd_{\mu}\phi_n^*
  \pd^{\mu}\phi_n-(m^2+\mu_n^2)|\phi_n|^2)- \int_{\K}dy\,
  V(\phi)\mid_{\phi=\sum\phi_n\psi_n}\right\}.
\end{equation}

To obtain eq. (\ref{d-p}) we used orthonormality of the spectrum
of $\Delta$. In the case of disconnected $\K$ the action is split
in independent non-interacting parts corresponding to each
connected component. Thus, without loss of generality from here on
we can assume that $\K$ is connected.

In what follows we will consider compactifications with small
compact size, in this case one can neglect $m^2$ in comparison
with nonzero $\mu^2_n$ in eq. (\ref{d-p}).

If one drops out the terms with nonzero $\mu_n^2$ one gets
classically compactified (or dimensionally reduced)
$D$-dimensional model. It is given by the following action,
\begin{equation}\label{reduced}
  S_{D}=\int_{\M}dx\left\{\frac12 (\pd_{\mu}\varphi
  \pd^{\mu}\varphi-m^2\varphi^2)-
  \frac{\bl}{4!}\varphi^4\right\},
\end{equation}
where $\varphi$ is the zero mode component of $\phi$, and coupling
$\bl$ is given by rescaling, $\bl=\lambda V_p^{-1}$,
$V_p=\int_{\K}dy$ is the volume of $\K$.

Action (\ref{reduced}) plays the role of the bare action which
will get corrections from KK fields in (\ref{d-p}). As we
mentioned in the Introduction, to be fully consequent we have to
add also the contribution of high energy modes of $\varphi$. The
last could be obtained by introducing an extra Pauli--Villars (PV)
regularisation for $\varphi$-loops (see e.g. \cite{sl-fadd}) with
cutoff mass $M\sim (V_p)^{-\frac{1}{p}}$  and computing one-loop
contribution from the regulator(s). For our purposes will suffice
three PV fields with the action,
\begin{equation}\label{pvaction}
  S_{PV}=\sum_{r=1}^{3}\int_{\M}dx\left\{\frac12 (\pd_{\mu}\phi_{r}
  \pd^{\mu}\phi_{r}-M_r^2\phi_{r}^2)-
  \frac{\bl}{4}\varphi^2\phi_{r}^2\right\},
\end{equation}
where the PV masses $M_r$ satisfy,
\begin{equation}\label{pvcond}
  \sum_{r}(-1)^r M_r^{2}=-m^{2}.
\end{equation}
Grassmann parity of $\phi_r$ is chosen to be $(-1)^{r}$.

The effective action we want to compute is defined as follows,
\begin{equation}\label{eff-act}
  e^{iS_{eff}(\varphi)}=\int d\phi_{M}
  \prod_{n\neq 0}d\phi_n\,e^{iS_{(D+p)}(\varphi,\phi_n)
  +iS_{PV}(\varphi,\phi_{r})}.
\end{equation}

The coupling $\lambda$ in (\ref{action}) gives rise to an infinite
number of couplings in compactified action (\ref{d-p}),
\begin{align}\label{poten}
  \frac{\lambda}{4!}\phi^4\rightarrow \frac{1}{4!}\lambda^{n_1\dots
  n_4}\phi_{n_1}\phi_{n_2}\phi_{n_3} \phi_{n_4} \\ \nonumber
  \lambda^{n_1\dots n_4}\equiv \lambda
  \int_{\K}dy\,\psi_{n_1}\psi_{n_2}\psi_{n_3}\psi_{n_4},
\end{align}
where, in general, $\lambda_{(l)}^{n_1\dots n_l}$ are all nonzero,
but for a smooth manifold $\K$ they decay faster than any power of
$n=\|n_1+\dots+n_4\|$ (where $\|\cdot\|$ is some properly defined
norm on the lattice $\Gamma$). We will not discuss further this
property, moreover in one-loop approximation it does not play any
r\^{o}le.

In fact, one-loop computations require only a small part of
couplings $\lambda^{n_1\dots n_l}$. Since for nonzero $\mu^2_n$
fields $\phi_n$ are non-propagating the one-loop diagrams contain
only interactions quadratic in KK modes $\phi_n$, $n\neq 0$,
\begin{equation}\label{vert}
  V_{(2)}=\frac{\bar{\lambda}}{4} \varphi^{2}|\phi_n|^2.
\end{equation}

To compute the one-loop effective action we also need the
propagators of the fields. They look as follows,
\begin{align}\label{propkk}
  D_{nn^*}^{-1} (p)=p^2 -\mu_n^2+i\epsilon, \qquad &\text{for
  $n\neq 0$},\\ \label{proppv}
  D^{-1}_{r}(p)=p^2 -M_r^2+i\epsilon, \qquad &\text{for PV regulator},
\end{align}
where $i\epsilon$ stands for the causal pole
prescription\footnote{An alternative choice would be Wick rotation
to Euclidean space-time as was used in \cite{eu}.}. Let us note
that both KK interaction term (\ref{vert}) and KK propagator
(\ref{propkk}) is of the same form as interaction term in
(\ref{pvaction}) and propagator for PV fields (\ref{proppv}). This
allows one to include KK and PV fields in the same set. From now
on let the index $n$  span both KK and PV fields.

Consider diagrams with KK/PV fields running in the loop and with
external $\varphi$ legs.

In what follows we use dimensional regularisation for KK-fields
and PV regulators. General one-loop diagram with $2N$ (truncated)
external legs has $N$ vertices and looks as follows,
\begin{multline}\label{diag}
  G(k_1,\dots,k_N)=\\
   \frac{\bl^N}{4^N}\sum_n\int
  \frac{d^{\tilde{D}}p}{(2\pi)^{\tilde{D}}}
  D_{nn^*}(p+k_1)\dots D_{nn^*}
  (p+k_1+\dots+k_N),
\end{multline}
where integration is performed over $\tilde{D}$-dimensional
Minkowski momentum space. As usual in dimensional regularisation
scheme coupling $\bl (\tilde{D})=\lambda_0 \kappa^{4-\tilde{D}}$,
where $\lambda_0$ is the dimensionless coupling and $\kappa$ is
the mass unity. Summation in (\ref{diag}) and on is performed
through the eigenvalue lattice of $\Delta$ where zero term is
substituted by PV regulator. Cutoff removing is obtained when
$\tilde{D}\to D$.

The formal divergence index of this diagram is $\omega_G=D-2N$.
This index, however, is computed termwise and does not reflect
possible divergencies due to the summation over $n$. From above it
follows the termwise divergencies described by $\omega_G$ are the
same for KK an PV modes. They are due to large momenta in low
dimensional directions and have low dimensional nature. From the
other hand, divergencies due to summation over KK modes correspond
to large momenta in compactified directions and, thus they have
the ``high-dimensional nature''.

Let us perform a change of the integration variable, $p\rightarrow
\frac{p}{\mu_n}$ for each $n$. This is legitimate since all
integrals are regularised.

After this the integral (\ref{diag}) looks as follows,
\begin{multline}
  G(k_1,\dots,k_N)=\\
  \frac{\bl^N}{4^N} \sum_n\left(\mu_n^2\right)^{-(N-\frac{\tilde{D}}{2})}
  \int \frac{d^{\tilde{D}}p}{(2\pi)^{\tilde{D}}}
  D(p+k_1^n)\dots D
  (p+k_1^n+\dots+k_N^n),\label{transfd}
\end{multline}
where $k_i^n=  \frac{k_i}{\mu_n}$.

One can see that all dependence on $n$ in integrand in
(\ref{transfd}) resides in $k_i^n$. All ``external momenta''
$k_i^n$ are proportional to a vanishing factor $\frac{1}{\mu_n}$.
So, one can safely expand Feynman integral (\ref{transfd}) in
powers of $k_i$. This yields,
\begin{multline}
   G(k_1,\dots,k_N)=\\
   \frac{\bl^N}{4^N}\left\{ I_{(0)}(\tilde{D},m)\sum_n\left(\mu_n^2\right)^{-(N-
  \frac{\tilde{D}}{2})}
  + I_{(2)}(k,\tilde{D})\sum_n
  \left(\mu_n^2\right)^{-(N+\frac{2-\tilde{D}}{2})}+\right.\\
  \dots+\left.I_{(r)}(k,\tilde{D})\sum_n
  \left(\mu_n^2\right)^{-(N+\frac{r-\tilde{D}}{2})}
  +\dots \right\},\label{expansion}
\end{multline}
where integrals $I_{(r)}$ are defined as follows
\begin{multline} \label{i's}
  I_{(r)}(\tilde{D})=
  \\ \left.\frac{1}{r!} q_{\mu_1}^{i_1}\dots q_{\mu_r}^{i_r}\frac{\pd^r}{\pd
  t_{\mu_1}^{i_1}
  \dots \pd t_{\mu_r}^{i_r}}
  \int \frac{d^{\tilde{D}}p}{(2\pi)^{\tilde{D}}}
  D(p+t_1)\dots D(p+t_N)\right|_{t_i=0},
\end{multline}
we introduced notations $q^i=\sum_{j=1}^{i} k_j$. Divergence
degree of each $I_{(r)}$ is $\omega_{G,r}=D-(2N+r)$. Let us note
that, starting from some $r\geq D-2N$ the integrals $I_{r}$ are
finite, therefore, expansion in $k_{i}^{n}$ is well defined.

One can see that expression (\ref{expansion}) has the standard
form of a high energy mode contribution, except for masses which
are substituted by the series over powers of $\mu_n^2$.

These series define well-known $\zeta$-function of the Laplace
operator, \cite{gilk},
\begin{equation}\label{zeta}
  \zeta_\Delta(s)=\tr(- \Delta)^{-s}={\sum_{n}}' (\mu_n^2)^{-s},
\end{equation}
where the prime means that summation is performed over the nonzero
eigenvalues $\mu_n^2$ (PV masses not included). These series are
convergent for $\R s>p/2$, but can be analytically continued to
other values of $s$.

Using definition (\ref{zeta}) one can rewrite eq.(\ref{expansion})
in the following form,
\begin{multline}
  G(k_1,\dots,k_N)=
  \frac{\bl}{4^N}\left\{I(\tilde{D})\left(\zeta_\Delta\left(N-
  \frac{\tilde{D}}{2}\right)+\sum_{r}(-1)^{r}M_{r}^{
  \tilde{D}-2N}\right)\right.\\
  +
  I_{(2)}(k,\tilde{D})
  \left(\zeta_\Delta\left(N+\frac{2-\tilde{D}}{2}\right)
  +\sum_{r}(-1)^rM_{r}^{\tilde{D}-2(N+1)}\right)+\dots \\
  \dots+\left.
  I_{(r)}(k,\tilde{D})\left(\zeta_\Delta
  \left(N+\frac{r-\tilde{D}}{2}\right)+\sum_{r}(-1)^r
  M_{r}^{\tilde{D}-(2N+r)}\right)
  +\dots \right\}.\label{withz}
\end{multline}

Expression in the r.h.s. of eq (\ref{withz}) can diverge in the
limit $\tilde{D}\rightarrow D$ due to the following two factors.
First one is singularity of some $I_{(r)}(\tilde{D})$ and second
one is singularity of $\zeta_\Delta(s)$.

Singularities of $I_{(r)}$ does not depend on index $n$.
Therefore, the renormalisability of the low dimensional reduced
model implies possibility to eliminate also singularities of this
type due to KK modes by a modification of already existing low
dimensional counter-terms. Let us note that $I_{(r)}$ is
proportional to the factor,
\begin{equation}\label{i}
  I_{(r)}(\tilde{D})\sim \Gamma\left(N+\frac{r-D}{2}\right),
\end{equation}
which is singular for $N+\frac{r-D}{2}=0,-1,\dots, \frac{D}{2}$,
when $D$ is even. At the same time
$\zeta_\Delta\left(N+\frac{r-D}{2}\right)$ in this points is
regular (see \cite{gilk}). Thus, the counterterms required to
cancel this type of singularities have the form
\begin{equation}\label{countertm}
  \Delta Z_{N,r}^{I}=\left(m^{2N+(r-D)}-\kappa^{2N+(r-D)}\zeta_\Delta
  \left(N+\frac{r-D}{2}\right)\right)\Delta Z_{N,r},
\end{equation}
where $\Delta Z_{N,r}$ is the low dimensional (reduced) model
counterterm and $\kappa$ is the ``mass unity'' of dimensional
regularisation.

Consider now divergencies due to singularities of
$\zeta$-function. It is known, that $\zeta_\Delta (s)$ has a
meromorphic extension to the entire complex plane except for
isolated simple poles on the real axis at $s=p/2-n$, $n=0,
1,\dots,\left[\frac{p}{2}\right]-1$. Residues of $\zeta_\Delta
(s)$ at this points can be expressed in terms of the heat kernel
invariants\footnote{We use notations of the Gilkey's book
\cite{gilk}.} $a_{2n} (\Delta)$ of the Laplace operator,
\begin{equation}\label{res}
  \Res_{s=\frac{p}{2}-n}\zeta_\Delta(s)=\frac{a_{2n}(\Delta)}
  {\Gamma\left(\frac{p}{2}-n\right)}.
\end{equation}

One can observe that $\zeta$-function variable is related to the
external divergence index as $s=-\frac{1}{2}\omega_{G,r}$. Thus,
eq. (\ref{res}) lead to singular contributions corresponding to
several diagrams with negative divergence index. In low
dimensional theory such diagrams are convergent and there are no
terms in the bar action (\ref{reduced}) to allow absorbtion of the
divergent part corresponding to such diagrams. In order to avoid
the non-renormalisability one have to get rid of such
contribution.

To do this consider invariants $a_{2n}(\Delta)$. They arise as
quotients in expansion of the diagonal part of the Heat Kernel at
$t=0$,
\begin{align}\label{hk}
  &K(t;y,y)=\sum_{n=0}^{\infty}t^{\frac{n-p}{2}}a_{n}(\Delta,y),\\
  &a_{n}(\Delta )=\int_{\K}dy\, a_{n}(\Delta,y).
\end{align}

For a Riemannian manifold they can be computed in terms of
integrals of local quantities (metric, spin connection and
curvature tensor). The first three nontrivial invariants look as
follows \cite{gilk},
\begin{align}\label{a's}\nonumber
  a_0(\Delta)&=\frac{V_p}{(4\pi)^{p/2}}, \\
  a_2(\Delta)&=-\frac{1}{6(4\pi)^{p/2}}\int_{\K}dy\, R, \\ \nonumber
  a_4(\Delta)&=\frac{1}{360(4\pi)^{p/2}} \int_{\K}dy\, (-12 \Delta
  R+5R^2-2R_{ij}{}^{i}{}_{k}R_{l}{}^{jlk}+2R_{ijkl}R^{ijkl}),\\
  \nonumber
\end{align}
where $R_{ijkl}$ and $R$ are respectively Riemannian and scalar
curvature. Invariants with odd index vanish, $a_{2n+1}=0$. Higher
invariants depend on higher powers of curvature and its covariant
derivatives.

Let define $\ell\equiv(V_{p})^{\frac{1}{p}}$ to be the
characteristic size of $\K$. From  the definition of
$\zeta$-function one can see that at some fixed point
$s=s_0$\footnote{From here on we drop the subscript $\Delta$ in
the notation of $\zeta$-function},
\begin{equation}\label{order}
  \zeta (s_0)\rightarrow L^{2s_0}\zeta (s_0)
\end{equation}
under global rescaling of the metric $g_{ij}\rightarrow
L^2g_{ij}$. Therefore, at $s_0$, $\zeta (s_0)\sim \ell^{2s_0}$.

If one takes the limit $\ell\rightarrow 0$ prior to cutoff
removing all Feynman diagram contributions with $\omega_{G,r}<0$
vanish. Since in this limit appears also divergent terms this
limit should be combined with cutoff removing in such a way that,
\begin{equation}\label{req}
  \lim_{
  \begin{array}{rl}
    \tilde{D}&\rightarrow D \\
    \ell&\rightarrow 0
  \end{array}}\zeta \left(N+\frac{r-\tilde{D}}{2}\right)=0,\qquad
  \text{for }N+\frac{r-D}{2}>0.
\end{equation}
Condition (\ref{req}) provides that limit $\ell\rightarrow 0$ is
reached faster than cutoff is removed which guarantee that after
the cutoff removing no term with $\omega_{G,r}<0$ will survive
(cfy. eq (\ref{withz})). Thus, the only contribution which
survives in this limit is given by the terms which renormalise the
bar action (\ref{reduced}).

From eq. (\ref{req}) it follows that cutoff removing
$\tilde{D}\rightarrow D$, and the limit $\ell\rightarrow 0$ must
combine in a way satisfying,
\begin{align}\label{lim}
  \tilde{D}&\rightarrow D, \\ \label{lim-even}
  \frac{(\kappa\ell)^2}{(D-\tilde{D})}&\rightarrow 0,
\end{align}
where $\kappa$ is the dimensional regularisation ``mass unity". In
what follows we will consider that $\ell$ decays as
\begin{equation}\label{ell}
  \ell^{2}\sim \kappa^{-2}\epsilon^{1+\alpha},\qquad 0<\alpha<1,
\end{equation}
where $\epsilon=4-\tilde{D}$. This is sufficient to satisfy
(\ref{lim-even}) without too fast decay of $\ell$.

Physically, such a limiting procedure means that the size of
compact dimensions should be \emph{below} the scale of low
dimensional effective quantum field theory or physically zero. In
other words, the quantum field theoretical approach is limited to
energies much smaller than the inverse size of compact dimensions.
This is required to make sense of compactified model as a quantum
field one.

As a result the effective theory is described by the same action
(\ref{reduced}) but with mass $m$ and coupling $\bar{\lambda}$
substituted by renormalised quantities. In general one can have
also the field renormalisation of $\varphi$, although in one-loop
approximation of $\phi^4$-model it is not the case.

Let us consider non-vanishing terms in more details, assuming that
$D=4$ and potential is given by (\ref{potential}). These terms
correspond to integrals with $\omega_{G,r}\geq 0$ and are
low-dimensionally UV divergent. As we already mentioned this terms
produce only renormalisation of the bar action (\ref{reduced}).
There are only two diagrams which contribute in one-loop approach
in our model. They are, respectively, two-point function,
\begin{equation}\label{two-point}
  G_{2}=-\frac{\bl}{4}\varphi^2 \sum_{n}\int \frac{d^{\tilde{D}}p}
  {(2\pi)^{\tilde{D}}}\frac{1}{p^2-\mu_n^2+i\epsilon}, \qquad
  \omega_{2}=2,
\end{equation}
and the zeroth term in $k$ expansion of $G_4(k)$,
\begin{equation}\label{four-point}
  G_{4,0}=\frac{\bl}{16}\varphi^4\int
  \frac{d^{\tilde{D}}p}{(2\pi)^{\tilde{D}}}
  \left(\frac{1}{p^2-\mu_n^2+i\epsilon}\right)^2,\qquad
  \omega_{4,0}=0.
\end{equation}
Integrals $G_2$ and $G_{4,0}$ lead respectively, to mass $m^2$ and
coupling $\bl$ renormalisation.

Computation of regularised $G_2$ and $G_{4,0}$ yields,
\begin{align}\label{g2}
  G_2&=-\frac{\bl}{4(2\sqrt{\pi})^{\tilde{D}}}
  \varphi^2\left(\zeta(1-\tilde{D}/2)+\sum_{r}(-1)^{r}
  M_{r}^{\tilde{D}-2}\right)
  \Gamma (1-\tilde{D}/2), \\
  G_{4,0}&=\frac{\bl^2}{16(2\sqrt{\pi})^{\tilde{D}}}\varphi^4
  \left(\zeta(2-\tilde{D}/2)+\sum_{r}(-1)^{r}
  M_{r}^{\tilde{D}-4}\right)
  \Gamma(2-\tilde{D}/2).
\end{align}

When $\ell$ and $2\epsilon=4-\tilde{D}$ are sent to zero according
to (\ref{lim}) and (\ref{lim-even}), functions $G_2$ and $G_4$
look like,
\begin{align}\label{div2}
  G_2&=
  \frac{\lambda_0}{4(2\sqrt{\pi})^4}
  (2\sqrt{\pi})^{2\epsilon}\times\\ \nonumber
  &\left(\ell^{-2}(\kappa\ell)^{2\epsilon}\zeta_0(-1+\epsilon)+\sum_{r}(-1)^{r}
  M_{r}^{2}(M_{r}/\kappa)^{-2\epsilon}\right)\Gamma(-1+\epsilon),\\ \label{div4}
  G_{4,0}&=\frac{\lambda_0^2}{4(2\sqrt{\pi})^4 }
  (2\sqrt{\pi})^{2\epsilon}
  \left((\kappa\ell)^{2\epsilon}\zeta_0(\epsilon)+
  \sum_{r}(-1)^{r}(M_{r}/\kappa)^{-2\epsilon}\right)\Gamma(\epsilon),
\end{align}
where the ``dimensionless" $\zeta$-function $\zeta_0 (s)$ is
defined as $\zeta$-function for the unity volume scaled manifold
$\K^{0}$, $\vol \K^{0}=1$, i.e. one obtained from $\K$ by
rescaling of the metric $g_{ij}^{0}=\ell^{-2}g_{ij}$. We used
(\ref{order}) to separate the dimensional factor from the
$\zeta$-function.

In the limit (\ref{lim}), (\ref{lim-even}) diagram $G_2$ produces
counterterm leading to renormalisation of the mass,
\begin{multline}\label{g2-div}
  \delta m^2\equiv\Delta Z_{2,0}=
  \frac{\lambda_0}{4(2\sqrt{\pi})^4}\ell^{-2}
  \left(-\frac{1}{\epsilon}(\zeta_0(-1)-m^2\ell^2)+\right.\\
  +((\zeta_0(-1)-\xi_{(2)})(-1+\gamma)-\zeta'_0(-1)-2\log
  (2\sqrt{\pi}\kappa\ell)\zeta_0(-1))\\
  +\left(\zeta_0(-1)\left(-\frac{1}{12}\pi^{2}-
  \frac{1}{2}\gamma^{2}-1+\gamma\right)\right.\\
  +(\xi_{(2)}+\zeta_0'(-1)+2\log(2\sqrt{\pi}\kappa\ell)
  \zeta_0(-1))(-1+\gamma)\\ \left.\left.
  -\frac{1}{2}\zeta_0''(-1)-2(\log(2\sqrt{\pi}\kappa\ell))^2
  \zeta_0(-1)-2\log(2\sqrt{\pi}\kappa\ell)\zeta_0(-1)\right)
  \epsilon\right)\\
  +\text{finite and vanishing terms},
\end{multline}
where $\gamma$ is the Euler constant, $\xi_{(2)}\equiv
2\ell^2\sum_{r}(-1)^r M_r^2\log M_r/\kappa$.

Analogously, $G_4$ leads to renormalisation of the coupling,
\begin{multline}\label{g4-div}
  \delta\bl\equiv\bl\Delta Z_{4,0}=\\
  \frac{\bl^2}{4(2\sqrt{\pi})^4}
  \left(\frac{1}{\epsilon}(-1+\zeta_0(0))+\xi_{(0)}+
  2\zeta_0(0)\log(2\sqrt{\pi}\kappa\ell)
  +\zeta_0'(0)+(1-\zeta_0(0))\gamma\right)\\
  +\text{vanishing terms},
\end{multline}
where $\xi_{(0)}=2\sum_{r}(-1)^r\log M_r/\kappa$.

As we expected the contribution of the KK and PV fields reduces in
the limit $\ell\rightarrow 0$ to mass and coupling renormalisation
which at first look seems trivial. In fact, at energies
$\sim\kappa\ll \ell^{-1}$ the low dimensional effective model we
obtained does not feel directly the fluctuations along $\K$,
however, as one can see from eqs. (\ref{g2-div}) and
(\ref{g4-div}) the coupling/mass renormalisation is sensible to
geometry of $\K$ through $\zeta$-function and $\ell$ present in
the counterterms. Thus, if the shape of $\K$ is varied the effect
of such variation consists in finite counterterms renormalising
respectively the coupling and the mass of propagating field
$\varphi$. This flow affects renormalised physical quantities as
well, provided fluctuations of $\varphi$ with momenta below
$\ell^{-1}$ are taken into consideration. This is because the
geometry dependent counterterms cannot be cancelled by constant
ones.

Flow of the parameters under the variation of the compact size
$\ell$ can be computed using renormalisation group methods. If one
renormalises the parameters at some value of the compact size
$\ell_0$ and than vary it one encounters some flow of the
renormalised parameters (renormalisation is $\ell$-independent),
$m_R=m_R(\ell)$ and $\bl_R=\bl_R(\ell)$. By dimensional arguments
it is clear that this flow is governed by the renormalisation
group. For example, for the renormalised coupling one can write,
\begin{equation}\label{rg1}
  -\ell \frac{\pd \bl_R}{\pd \ell}=\beta(\bl_R),
\end{equation}
where $\beta(\bl)$ is the Calan-Simanzik beta function, which in
the one-loop approximation is given by
\begin{equation}\label{beta}
  \beta(\bl)=\kappa\bl\frac{\pd Z_{4,0}}{\pd \kappa}=
  \frac{\bl^2}{2(2\sqrt{\pi})^4}(\zeta(0)-1).
\end{equation}
here we used that at $s=0$, $\zeta_0(0)=\zeta (0)$.

Solving eq. (\ref{rg1}) yields for the coupling flow,
\begin{equation}\label{phi4}
  \lambda(\ell)=\frac{\lambda(\ell_0)}{1+(a_p-1)
  \frac{\lambda(\ell_0)}{32\pi^2}\log\frac{\ell}{\ell_0}},
\end{equation}
where $\lambda(\ell)$ and $\lambda(\ell_0)$ are renormalised
couplings computed when the size of $\K$ is, respectively, $\ell$
and $\ell_0$, we also used eq. (\ref{res}) to substitute
$\zeta(0)$ by its value $a_p (\Delta)$

\section*{Discussions}
In the actual paper we considered compactifications of
high-dimensional quantum field models.

We have shown, that a renormalisable compactified model can be
defined when the size of compact dimensions is of the order of
cutoff scale of the low dimensional model. In dimensional
regularisation scheme this corresponds to vanishing of the size of
compact dimensions when the cut off dimension $\tilde{D}$
approaches the physical value $D$.

In condition (\ref{lim}), (\ref{lim-even}) we required that the
compact size decays faster than cutoff is removed. This condition
arose from the requirement not to have nonrenormalisable terms in
the low dimensional effective theory. In fact, one can try to
slightly relax this condition by requiring,
\begin{equation}\label{soft}
  \frac{(\kappa\ell)^2}{\epsilon}\rightarrow \text{finite},
\end{equation}
rather than zero. Our analysis in this case still remains
applicable, but resulting model will not be a renormalisable one.

Indeed, condition (\ref{soft}) still leads to absence of
nonrenormalisable infinite contributions to the effective action,
but in this case beyond the mass/coupling renormalisation one
would have new finite terms in the effective action of the
following types $(\pd\varphi^2)^2$ and $\varphi^6$ (in general
also terms like $(\pd^2\varphi)^2$ may appear, however in one-loop
they are absent). These are, so called, IR irrelevant operators
so, if the energy scale is lowered they are believed to decouple
then one recovers the old situation. Due to the presence of higher
derivative terms this this non-renormalisability could be
interpreted as effective switch-on of additional degrees of
freedom from the compact dimensions.

Having our results at hand one can draw the following qualitative
physical picture. At energies much lower than the size of compact
dimensions one has a renormalisable low dimensional QFT whose
parameters depend on the geometry of compact dimensions. When the
energy is increased new degrees of freedom appear, due to the
non-renormalisable interactions and higher order terms at energy
$\kappa\sim\ell^{-1}$.

An alternative approach to define a renormalisable compact model
may consist in compactifications to non-commutative compact
manifolds \cite{connes}. Fields on non-commutative spaces
\cite{noncom} are now under intensive study in connection with
matrix models, \cite{cds}.

So far, we considered the scalar field model. However,
generalisation to other models both ordinary and supersymmetric is
more or less straightforward.

In fact one can consider more general compactifications, when
manifold $\W$ is a fibre bundle $\W=\M\rtimes\K$, where Minkowski
space $\M$ is the base and $\K$ is the fibre. In actual paper one
may consider that compactification is of the later type but the
fiber $\K$ depends on Minkowski space time-point adiabatically.
The last means that for large enough space-time regions this fibre
bundle should look almost as a direct product.

Before concluding let us make a few comments. First, let us note
that coupling flow (\ref{phi4}) differs from the standard flow one
would have in the $\phi^4$ model by a factor $(1-a_p)$ in the
denominator. This factor in general changes the rate of the flow
and in the case when $a_p\leq 1$ even its character. For $p=2,4$
e.g. the respective $a_p$ are given by second and third eqs.
(\ref{a's}).

In actual paper we neglected contributions of the order
$O(m^2\ell^2)$. These can be taken into consideration by
substituting $\Delta$ by $\Delta+m^2$ in eq. (\ref{spec1}) and
(\ref{spec2}), which results in using $\zeta$-function of
$\Delta+m^2$ rather than one of $\Delta$. All analysis in this
case remain the same except a slight modification of heat kernel
invariants (\ref{a's}).

Renormalisation group flow (\ref{phi4}) makes sense only when
theory is well-defined also beyond the one loop. Although  our
analysis heavily used specific one-loop tools, such as
Pauli--Villars regularisation and zeta function sums this analysis
can be extended beyond one-loop by some technical modifications,
and, we believe, the present conclusions to remain valid.

\subsection*{Acknowledgements}
I would like to thank A.A.~Slavnov, D.~Maison, K.~Zarembo,
H.J.W.~M\"uller-Kirsten, R.~Manvelyan, K.~Sailer for useful
discussions.

This work was completed during my stay at the Theoretical Physics
Department of Kossuth Lajos University in Debrecen and the
Department of Theoretical Physics, Universit\"at Kaiserslautern.

\end{document}